\documentstyle[12pt]{article}
\newcommand{\xsubsection}{\subsection}

\newcommand{\BR}{{\bf R}}
\newcommand{\BC}{{\bf C}}

\newcommand{\BZ}{{\bf Z}}
\newcommand{\CL}{{\cal L}}
\newcommand{\CN}{{\cal N}}
\newcommand{\Tr}{{\rm Tr}}
\newcommand{\tr}{{\rm tr}}
\newcommand{\Dslash}{{\rlap{\hskip0.2em/}D}}

\newcommand{\I}{I}
\newcommand{\II}{\relax{I\kern-.10em I}}
\newcommand{\IIa}{{\II}a}
\newcommand{\IIb}{{\II}b}
\newcommand{\ap}{\alpha'}
\newcommand{\sqap}{\sqrt{\alpha'}}

\def\vev#1{\langle{#1}\rangle}
\newtheorem{claim}{Claim}
\begin{document}
\begin{titlepage}
\begin{flushright}
RU-96-91\\
hep-th/9610041
\end{flushright}
\vfil
\begin{center}
{\large\bf Superstring dualities, Dirichlet branes and the \\
small scale structure of space} \\
\vskip 1cm
Michael R. Douglas\footnote{Partially supported by
DOE grant DE-FG05-90ER40559, NSF PHY-9157016 and the A. P. Sloan Foundation.}\\
Dept. of Physics and Astronomy\\
Rutgers University\\
Piscataway, NJ 08855-0849, USA\\
{\tt mrd@physics.rutgers.edu}\\
\end{center}
\vskip 0.5cm
\begin{abstract}
We give a broad overview of superstring
duality, Dirichlet branes, and some implications of both for questions
about the structure of space-time at short distances.
\end{abstract}
\vskip 0.5cm
\begin{center}
{\it To appear in the proceedings of the LXIV Les Houches
session on `Quantum Symmetries', August 1995, eds.
A. Connes and K. Gaw\c{e}dzki.}
\end{center}
\vfil
September 1996
\end{titlepage}

\section{Introduction}
Superstring theory has been studied intensively since 1984,
when the discovery by Green and Schwarz of anomaly cancellation
convinced many physicists
that it provides a consistent theory of perturbative quantum
gravity, gauge interactions and chiral matter.
The basic difficulties in quantizing general relativity and
supergravity (non-renormalizability or at least strong coupling at the
Planck scale) are visible at low order in the loop expansion, while
superstring theory was shown to be well-defined and
finite to all orders.

Although this was quite an achievement, and much has been understood
at this level, it has been clear for some time
that to answer the questions of direct physical interest,
one needs non-perturbative results.
These questions fall into two broad areas.

The first question is how does the standard model of particle physics
emerge at low energies, those far below the natural scales of string theory
and quantum gravity.  All plausible scenarios for this so far involve
supersymmetry at low energy, and superstring theory leads to
reasonable supersymmetric extensions of the
standard model given remarkably few assumptions,
as explained in \cite{GSW}.
Supersymmetry must then be broken by non-perturbative effects,
and detailed predictions depend crucially on this physics.

The second question is to understand
situations involving strong gravitational
effects.  The most famous examples are the
by now classic problems of black hole physics, such as what happens to a
black hole at the end point of Hawking radiation, and in particular is
this process governed by unitary quantum evolution.
Many simpler situations exist as well.
For example, consider a limit in which the volume of a
non-contractible cycle in an internal space (e.g. a Calabi-Yau)
goes to zero, and its curvature diverges.
It has long been thought that our conventional ideas of space-time
must be modified in such extreme situations and at very short distances.

Over the last two years, revolutionary progress in supersymmetric
field theory and superstring
theory has brought some of these questions within reach.
Many of these developments were described by Dijkgraaf,
Greene and other lecturers at this school.
At the time of the school, the subject was evolving
extremely rapidly, and the developments of the next few months
greatly extended the reach of the new ideas and significantly clarified them.
Thus, rather than artificially limit the discussion to the perspective
of summer 1995, it appeared to me to be far more useful to
adopt a later perspective
(summer 1996) while addressing some of the themes which came up in
lectures and discussions at and inspired by the school.

The present contribution is a broad overview of superstring
duality, Dirichlet branes, and the implications of both for questions
about the structure of space-time at short distances.
It is intended to be readable by physicists
and mathematicians without a detailed knowledge of string theory,
and serve as an introduction to more extensive reviews such as
\cite{Schwarz,Sen} for superstring
duality, \cite{CJP,PolRev} for Dirichlet branes and
\cite{Duff} for M theory.
(Another overview is \cite{PolCol}.)
Space did not permit
mentioning many of the important developments; in particular we almost
completely omit the topic of duality in compactified string theories.
Many of the concepts were discovered in that context and thus we will
not attempt to describe the history but again refer to these reviews.
(A few of the papers which must be mentioned are
\cite{HullTown,SchSen,Strom,Wit1}).

After a brief discussion of field theoretic duality,
we describe the dualities
between the five ten-dimensional superstring theories and
eleven-dimensional ``M theory.''
We then give an introduction to Dirichlet
branes (D-branes),
a particularly simple class of soliton in superstring theory,
which enters into and simplifies many arguments in duality.
We then give an overview of the works \cite{DFS,DKPS,kp,Shenker},
which showed why Dirichlet branes are particularly relevant
in studying the nature of space-time at short distances,
and combine this with a result of \cite{dm} --
D-branes can see non-trivial topology
in a novel way -- to see that the topology and geometry relevant for
general relativity must be embedded in a larger,
essentially non-commutative structure in string theory.
Finally, we discuss a few of the many open questions.

\section{Duality and Solitons in Supersymmetric field theory}

The central ideas behind the new developments
are strong-weak coupling duality in abelian
gauge theories, and
precise bounds on the masses of ``BPS states'' in supersymmetric theories.
Gauge theories and string theories have a rich spectrum of solitonic states,
both particles and ``$p$-branes,'' objects extended in $p$ spatial dimensions.
Classically, their masses and tensions typically behave as $m \sim 1/g^2$
in terms of a conventional three-point coupling $g$.  In certain
string theories (as we will discuss), one finds solitons
with $m \sim 1/g$.
Some of these, the BPS states,
belong to reduced multiplets of supersymmetry, and for them,
the classical mass formula is valid quantum mechanically as well.\footnote{
This will be true in the cases we discuss, which have enough supersymmetry
to protect the low energy effective Lagrangian from quantum corrections.
The general statement is that the classical mass formula derived from the
exact low energy effective Lagrangian is valid.  This is an equally powerful
statement which (for example) was central in Seiberg and Witten's solution
of $\CN=2$ super Yang-Mills theory.}
Thus, as $g$ becomes large, the solitons become the lightest states and
dominate the dynamics of the theory.

The simplest example of how this statement can be made precise
is the ``proof'' of duality
for four-dimensional $\CN=4$ super Yang-Mills theory
(henceforth, SYM).  This theory has
a moduli space of vacua and at generic points the gauge symmetry is broken
to the Cartan subalgebra of the gauge group.  The other generators of the
group correspond to massive gauge bosons, with mass $m$ set by the scale
of gauge symmetry breaking, and these are the lightest charged states in
the theory.  If we work at energies far below $m$, the theory is effectively
the same as abelian gauge theory, a non-interacting theory.
(All of these points will reappear in our discussion of D-branes
in section 4, with a rather different physical picture.)

The BPS solitons are magnetic monopoles with mass $m/g^2$, and
for $g>1$, they become the lightest charged states.  After a duality
transformation exchanging electric and magnetic charge, we get a weakly
coupled gauge theory with these charged particles.  Now -- and this is the
main point where we use the fact that this is $\CN=4$ SYM -- the monopoles form
supersymmetry multiplets containing spin $1$ particles
which are identical to the original gauge boson multiplets.  We only know
of one sensible field theory containing charged spin $1$ particles,
gauge theory.  Thus, even without having a precise change of variables
from the original Yang-Mills theory to the dual theory, we can assert that
for sufficiently large $g$, the dual theory must be a gauge theory.
Finally, $\CN=4$ SYM is the unique gauge theory with these degrees of freedom
and symmetry.
The final result, a dual theory isomorphic to the original
theory at weak coupling, is special to this example, but the line
of reasoning is general.

All this was Montonen and Olive's original argument for their conjecture
of duality \cite{MontOl}
and it is obviously far from a mathematical or even physicist's
proof.  However many physicists are convinced by the argument, not just
by the beauty of the result and the large framework it has been fit into,
but because upon reflection, it is hard to come up with a way for it to fail
which would not be even more surprising than the duality conjecture.

The first step in a proof would be to show that $\CN=4$ SYM
actually exists (there is no `supersymmetry anomaly').
It is unlikely that there is any
regulated form of the theory with the supersymmetry (besides string theory!)
and nobody knows how to even begin to prove this, but in every known
anomalous physical theory (one in which a classical symmetry is not present
in the quantum theory), there is a perturbative or semiclassical
computation which
exhibits the anomaly.  These computations have been done for $\CN=4$
SYM and no anomaly is seen.

One might imagine that some non-BPS states of which we are unaware
are the true lightest states.  They would have to be uncharged (the BPS
mass is a lower bound for charged states) and this forces their couplings
to the abelian gauge fields to be non-renormalizable and
(for generic parameters) less important
at low energies than the minimal gauge theory couplings.
Although these points speak against this possibility, it has not been shown
to be inconsistent, and this is perhaps
the most reasonable way the duality hypothesis could fail.
However, $\CN=4$ supersymmetry is a strong constraint and
no sensible scenario of this sort has been proposed.

Then, assuming the conventional principles of field theory apply
at low energies (since the theory is essentially Maxwell's theory in
this regime, this is very well motivated), the remaining element of the
argument which is not straightforward to check is the assertion that the
only sensible field theories containing particles with spin $1$ (and no higher)
are gauge theories.
Although this is not easy to prove, it has been a central point in theoretical
physics for some time (it is the foundation of the theoretical justification
of the Standard Model) and almost all physicists are convinced of it.

This physical background may help explain why,
once the idea of duality was taken
seriously, the remarkable paradigm shift of 1994-1995 was accepted
so quickly.
We can add to this the impact of Seiberg's work on supersymmetric
gauge theory and especially the celebrated
Seiberg-Witten solution of $\CN=2$ SYM, which after twenty years of effort
gave the first analytic results on confinement in four-dimensional gauge
theory, and demonstrated that these ideas could solve problems
of universally accepted physical importance \cite{IntSei}.

\section{Duality and Solitons in Superstring theory}

We begin by reviewing the field content and low energy effective
Lagrangians of superstring theories in ten dimensions.
The particle content is found by quantizing a single superstring,
and each massless particle is associated with a field in the effective
Lagrangian.
This analysis is described in many references such as \cite{GSW}.
Although it is too long a story to repeat here, we outline
it before quoting the Lagrangians,
to make the point that their field content, which at first sight
may seem somewhat arbitrary,
has a simple origin whose essence can be summarized fairly
briefly.
Given these fields, it has been shown
that in each case there is a unique low energy Lagrangian \cite{susy}.

\xsubsection{Spectrum of superstring theories}
The motion of a superstring world-sheet is governed by a two-dimensional
quantum field theory, whose degrees of freedom always include a map
embedding the world-sheet into space-time.
A specific theory is characterized by the additional
degrees of freedom which appear.
In the type \IIa\ and \IIb\ closed string theories, these are
two ten-dimensional spinors $\theta^1$ and $\theta^2$,
giving rise to $\CN=2$ supersymmetry.
The two cases are distinguished by
their relative chiralities -- opposite for \IIa\ (so the theory has a parity
symmetry) and the same for \IIb.
The heterotic strings are closed strings with $\CN=1$ supersymmetry
and thus a single ten-dimensional right moving
spinor $\theta^1$, and $32$ left moving fermions
$\lambda^I$.  These are singlets under Lorentz symmetry but admit an obvious
action of $SO(32)$.  The full definition of the theory specifies a sum over
sectors with various twisted boundary conditions, distinguishing
theories with $SO(32)$ or $E_8\times E_8$ gauge symmetry.

The type \I\ string contains open strings,
and a subset of the type \IIb\ closed strings as well.
The boundary conditions of an open string
relate $\theta^1$ to $\theta^2$ and the result is that the pair behaves
as a single ten-dimensional spinor degree of freedom, and the theory has
only $\CN=1$ supersymmetry.
Furthermore, the existence of the
boundaries allows us to introduce a discrete
degree of freedom or ``Chan-Paton factor'' at each
boundary.
Open string states are labelled by a pair of such choices,
say $1\le I,J\le N$.
We could notate a field creating an open string with a specified
embedding $X(\sigma)$ and auxiliary degrees of freedom as
$\phi^I_J[X(\sigma),\theta(\sigma)]$.

The basic open string interaction is a joining of strings at the endpoints,
and if we make the natural definition that the Chan-Paton factors must agree,
it will include a matrix multiplication $\sum_J\phi^I_J\phi^J_K$.
Thus open strings come with a natural algebraic structure, from which
arises the Lie algebra structure required for non-abelian gauge
symmetry.\footnote{The seminar actually given at Les Houches
involved a lengthy discussion of
extensions of this algebraic structure to the entire space of loop functionals,
provided by Witten's string field theory \cite{WitSFT}
and by matrix models \cite{matmod}.
%This was inspired by Connes' construction of ...
It is a mark of how deeply the recent developments
cut that the fundamental role of such constructions has
become completely unclear.  The strongest argument against their fundamental
role is the existence of limits of the
theory which do not contain fundamental strings.}
At the quantum level, consistency requires the gauge group to be $SO(32)$.
This comes from the choice $N=32$ and the identification
$\phi^I_J[X(\sigma),\theta(\sigma)]=\phi^J_I[X(-\sigma),\theta(-\sigma)]$,
which expresses the identification of the open string with
its orientation reversal, and makes the vector potential antisymmetric.
These conditions are forced on
us only in the particular case of open strings in ten dimensions.

The essential step in finding the massless spectrum is to identify
the zero modes of the additional world-sheet
degrees of freedom, as the massless
one particle states must form a representation of their algebra.
The result for the Lorentz non-singlet fermionic zero modes (those
carrying spin) is easily stated -- there is one for each
supercharge which acts non-trivially on the state.
More explicitly, a massless particle state admits
the little group symmetry $SO(8)\subset SO(9,1)$, and each ten-dimensional
world-sheet spinor contributes a zero mode
$p_\mu \gamma^\mu \theta=0$, a spinor of $SO(8)$.
Each has eight real components and
thus the massless states of the type \II\
strings admit an action of a Clifford algebra with $16$ generators and come in
a $256$ component multiplet of supersymmetry, while the massless states
of the type \I\ and heterotic strings admit an action
of a Clifford algebra with $8$
generators and come in multiplets with at least $16$ components.
The Lorentz representations of the states are thus determined and it is a short
exercise to show both that a graviton is present and --
the non-trivial
result we will use from this analysis -- that
the type \II\ multiplets contain an two-form gauge potential
and additional bosonic states in
the bispinor of $SO(8)$.  The bispinor is equivalent to the direct
sum of antisymmetric forms which appear in the effective Lagrangians.

\xsubsection{Effective Lagrangians for superstring theories}
%We proceed to quote the massless field content of the various theories.
The type \II\ strings have $\CN=2$, $d=10$ supersymmetry and thus a metric
$g_{\mu\nu}$,
two ten-dimensional gravitinos, and additional spinors to fill out $128$
fermionic states.  The additional bosonic fields are the scalar `dilaton'
$\phi$, a differential $2$-form gauge potential $B^{(2)}$,
and a sum of odd rank (for \IIa) or even rank (for \IIb)
$p+1$-form gauge potentials $C^{(p+1)}$.\footnote{
We denote the rank of a potential as $p+1$ for reasons to be explained shortly.
$B^{(2)}$ is often called the ``Neveu-Schwarz'' or NS two-form, and
the $C^{(p+1)}$ are called ``Ramond-Ramond'' or RR forms,
from their origins in the superstring spectrum.}
The bosonic part of the type \II\ effective action is then\footnote{
The actions (\ref{typeII}) and (\ref{sgeleven})
also contain cubic Chern-Simons couplings, schematically
$\int B\wedge dC\wedge dC$, but
they are not needed to verify the statements made here.
}
\begin{equation}\label{typeII}
\CL_\II = \int d^{10}x\ e^{-2\phi}
\left(\sqrt{g}R + 4||d\phi||^2 + ||dB^{(2)}||^2\right)
+ \sum_{p=0,2\ \IIa\atop p=-1,1,3\ \IIb} ||dC^{(p+1)}||^2 .
\end{equation}
The heterotic theory contains a subset of these fields as well as
non-abelian gauge fields with field strength $F$.  Its
bosonic effective action is
\begin{equation}\label{het}
\CL_{het} = \int d^{10}x\ e^{-2\phi}
\left(\sqrt{g}R + 4||d\phi||^2 + ||dB^{(2)} - \omega_3||^2
+ {1\over 4}\Tr ||F||^2 \right)
\end{equation}
where $\omega_3$ is a sum of Chern-Simons three-forms constructed from
the spin (Lorentz) connection and the non-abelian gauge connection \cite{GSW}.
The type \I\ theory contains a subset of the type \IIb\
fields, and non-abelian gauge fields $F$ from the open strings:
\begin{equation}\label{typeI}
\CL_\I = \int d^{10}x\ e^{-2\phi}
\left(\sqrt{g}R + 4||d\phi||^2 \right)
+ e^{-\phi}~\Tr ||F||^2 + ||dC^{(2)}- \omega_3||^2 .
\end{equation}
The type \II\ effective Lagrangians have gauge symmetries
$\delta B^{(2)} = d\Lambda^{(1)}$ and $\delta C^{(p+1)} = d\lambda^{(p)}$.
The standard gauge transformation laws are modified in the heterotic and
type \I\ theories, as is clear by the presence of the term $\omega_3$.
The full story is a bit subtle and involves a cancellation of gauge anomalies
after quantization (the Green-Schwarz mechanism \cite{GSW}).

We did not write an overall coupling constant $1/\hbar$ or $1/G_N$
in front of the Lagrangians, because it can be absorbed by a shift
of the dilaton $\phi$
and field redefinitions.  However, there is a moduli space of vacua
characterized by the expectation value $\vev{\phi}$ which will control
an effective coupling constant
$g_s=e^{\vev{\phi}}$.  The presence of this scalar also raises
an important point: there is a family of metrics (with parameter $a$)
\begin{equation}\label{Weyl}
G^{(a)}_{\mu\nu} \equiv e^{a\phi} g_{\mu\nu}
\end{equation}
several of which naturally appear in the discussion.
Lengths and masses must be quoted with
respect to a particular metric or `frame.'
The only one with an obvious
preferred status in the low energy Lagrangian is the Einstein metric,
the one which is governed by the Einstein action
$\int d^{10}x \sqrt{G}R[G]$.
However, this is not
the natural metric when a preferred (weakly coupled) BPS state is present,
as the Einstein mass of the preferred state will
usually depend on the coupling constant.
The Lagrangians above are written using the ``string metric,''
in terms of which the tension $T$ of a fundamental
string is independent of the coupling.
We implicitly set it to $1$ above;
when we want to make it explicit we use
the standard constant $\ap\equiv 1/T$ of dimensions $({\rm length})^2$
instead.
\iffalse
In this metric, the topological expansion
of fundamental string perturbation theory is controlled by the
string coupling constant $g_s=e^\phi$, and a diagram of Euler character
$\chi$ appears at order $e^{-\chi\phi}$.
The leading interactions of closed strings come from the spherical topology
with an incoming `tube' for each closed string, corresponding to
$e^{-2\phi}=1/g_s^2$, and those of open strings from a disk topology with
incoming `strips,' corresponding to $e^{-\phi}=1/g_s$.
\fi

To find the strong coupling limit of
the theories, we begin by classifying BPS solitons.
The result is easy to state:
\begin{claim}
For every $p+1$-form gauge potential,
there is a unique associated electrically charged BPS $p$-brane,
and a unique associated magnetically charged BPS $(6-p)$-brane.
\end{claim}
%[ what about gauge and symmetric five-brane? ]
The associated topological quantities are just the electric
and magnetic charges $Q_E = \int *dC$ and $Q_M = \int dC$ respectively.
Thus we must explain what a $p$-brane is before proceeding.

\xsubsection{$p$-branes}
A $p$-brane is a solution of the classical equations of motion,
independent of $p+1$ coordinates $y^\mu$ and localized in the remaining
(here $d=9-p$) coordinates, say around $x^i=0$.
We write the solution schematically as $\phi(x)$.
Such solutions are often determined by the solution in a subsector with
a topologically non-trivial field configuration.  For example,
a self-dual Yang-Mills configuration gives rise to a solution of
Einstein-Yang-Mills (still depending on four coordinates),
with metric non-trivial but determined by the gauge field.
All BPS solitons have this topological character.

Solitonic solutions generally have continuous parameters
(moduli or collective coordinates).
Any solution which breaks translation invariance in $d$ dimensions will
have moduli $\phi(x;X)=\phi(x+X;0)$.
Others might be predicted by symmetry or appear non-generically --
for example, the instanton
has a scale size and an embedding in the gauge group.

Configurations with slowly varying moduli are approximate solutions.
Choosing local coordinates $m$ on the moduli space such that
$\phi(x;m)$ is the solution with moduli $m$,
the nearby configuration $\phi(x;m(y))$ will have action proportional
to $\int d^{p+1}y\ (\partial m)^2$, so the $p$-brane admits
fluctuations with arbitrarily small action.

Now we can regard the choice of $p+1$-dimensional hyperplane (or
``world-volume'') and
map $m(y)$ from the world-volume to moduli space as specifying the state
of the $p$-brane.  The $m(y)$ behave as local fields on the world-volume,
and a $p$-brane in isolation will be governed by an effective action
derived by substituting the ansatz
$\phi(x;m(y,t))$ into the original ten-dimensional action.
This action will include
the metric on moduli space, its supersymmetrization, couplings to other
bulk fields, and so forth.
Its low energy limit is largely determined by the symmetries
broken by the soliton.

The universal part of the effective action governs the transverse
fluctuations of the $p$-brane.
It is determined by the charge $Q_p$ and the
original Poincar\'e invariance to be
the Nambu action with a minimal gauge coupling,
\begin{equation}\label{braneact}
S = \int T_p f^*(V) + Q_p f^*(C^{(p+1)}).
\end{equation}
The tension $T_p$ is proportional to $Q_p$ with ratio a function
of the vacuum moduli (here $\vev{\phi}$ or $g_s$) fixed by the
BPS condition in the ten-dimensional field theory.
$f:\BR^{p+1}\rightarrow \BR^{10}$ is the embedding,
$f^*(V)$ is the volume element induced on the brane from the space-time
metric $G_{\mu\nu}$, and $f^*(C^{(p+1)})$ is
the pullback of the gauge potential.
The other zero modes will have kinetic terms and
interactions consistent with symmetry but dependent on the particular case.

To return to the example of the self-dual gauge field, it can
be promoted to a solution of ten dimensional superstring theory
which is localized in four and extended in six dimensions.
This is a five-brane, an extended object filling five dimensions
of space as well as time.
Its non-zero action $\int \tr F^2=4\pi^2$ and the coupling dependence
in the Lagrangian determines its tension to be $T_{5h}=4\pi^2/g_s^2$
in the heterotic string, and $T_{5\I}=4\pi^2/g_s$ in the type \I\ string.

For $p=1$, (\ref{braneact})
is a form of the superstring action we might have used
to begin a detailed discussion of superstring quantization.
Thus a string soliton is identical to a fundamental string
at this level.  They differ in their additional interactions.
A fundamental string (by definition) admits a limit (zero string coupling)
in which all embeddings of its world-volume are possible, with
precisely the action (\ref{braneact}).  For the string soliton
(\ref{braneact}) is an approximation which will break down (for example)
when the world-sheet self-intersects.  The basic hypothesis underlying
superstring duality is that, despite this difference in our
explicit definitions of the objects, they are continuously connected
by varying the vacuum moduli.
In a limit in which $T_1$ for a particular soliton string becomes much smaller
than other dimensionful quantities in the theory,
that string will become a fundamental
string.  This hypothesis has the same character as the
arguments of section 2 but
is beyond explicit verification, even for a single string.  Nevertheless
it is amply justified by the results.

Having explained what a $p$-brane is, we return to describing the BPS
$p$-branes of the various string theories.
First, the fundamental string gives rise to a BPS state,
electrically charged under $B^{(2)}$.
Note that there must be an $S^7$ surrounding the object to have a non-zero
charge and thus only a fundamental string infinitely extended
in space will actually
be a BPS state; states of the original quantized loops of string are not BPS.
Note also that an infinitely extended fundamental
string in the type \I\ theory is not BPS.  It is not even a stable state
-- it can break into open strings.

To prove the claim,
we also need to find the magnetic $B^{(2)}$ (or NS) five-brane,
as well as $p$-branes
charged under the gauge fields $C^{(p+1)}$.
The coupling involving $\omega_3$ can be used to show
that in the heterotic string,
the NS five-brane is just the five-brane associated with self-dual gauge
fields.  An NS five-brane
also exists in type \II\ as a conventional soliton solution.

The $p$-branes associated with the RR gauge fields $C^{(p+1)}$ can be found
as solitons, and are equivalent to the D-branes described in the
next section.  In type \I\ theory the five-brane associated with
self-dual gauge fields is one of these.
They have some unusual characteristics following from the mixed
dilaton dependence
of the action -- notably, their tension has the unusual coupling
dependence $T_p \sim 1/g_s$.

We can now determine the object which governs the strong coupling
limit of each theory, using ``Hull's criterion.'' \cite{Hull}
We would like to find the lightest of the various $p$-branes
as $g\rightarrow\infty$, but
we cannot simply compare tensions $T_p$ for different $p$ as these
quantities have different dimension.
In our examples this will not be a serious problem, but in compactification to
lower dimensions many objects appear (branes wrapped around homology cycles)
and it is less obvious which object dominates.
Hull's criterion states that it will be the
object of smallest $m=(T_p)^{1/(p+1)}$.
Branes of high $p$ are strongly disfavored by this criterion, and by now
there is good evidence that they do not play the role of fundamental objects.

We proceed to consider the cases.
\xsubsection{Type \IIa}
Type \IIa\ superstring theory has a particle (zero-brane) with
$m\sim 1/g_s\sqrt{\alpha'}$ and a membrane (two-brane) with
$T_2\sim 1/g_s(\alpha')^{3/2}$.  The strong coupling limit
is controlled by the zero-brane.

This fact was used by Witten \cite{Wit1}
to argue that the strong coupling limit
of type \IIa\ superstring theory is in fact an eleven-dimensional theory
whose low energy limit is eleven-dimensional supergravity, but with space-time
the manifold $\BR^{10}\times S^1$.  Eleven-dimensional supergravity has
a metric, gravitino and a single $3$-form potential, and
the bosonic Lagrangian
\begin{equation}\label{sgeleven}
\CL_{11} = {1\over l_p^9}\int d^{11}x\ \sqrt{g}R + ||dC^{(3)}||^2 .
\end{equation}
The theory has no dimensionless coupling constant; the
effective coupling constant is $l_p E$, energy measured in Planck units.

It was long known that $S^1$ (Kaluza-Klein)
dimensional reduction of (\ref{sgeleven}) followed by a metric substitution
(\ref{Weyl}) produced the \IIa\ low
energy Lagrangian (\ref{typeII}).  The $g_{11,11}$ component
of the metric (determining the radius $R_{11}$ of the $S^1$)
becomes the ten-dimensional coupling, leading to the
relation $R_{11} = g_s^{1/3}\sqap$.
The Kaluza-Klein gauge field $g_{11,\mu}$ becomes $C^{(1)}_\mu$,
the vector potential coupling to the zero-brane.
Thus the ten-dimensional zero-brane charge is reinterpreted as momentum in the
eleventh dimension, $P_{11}$.  This momentum is quantized,
$P_{11}=n/R_{11}$, and the BPS argument guarantees that
the ten-dimensional mass $m$ of an eleven-dimensional massless state is exactly
$P_{11}$.  The vanishing $m = 1/g_s$ of the zero-brane mass
in the strong coupling limit is interpreted as the limit
$R_{11}\rightarrow\infty$.

Eleven-dimensional supergravity is neither finite nor renormalizable,
so its connection with superstring theory provides the first convincing
argument that it indeed exists as the low energy limit of a well-defined
quantum theory.  This theory has been dubbed ``M theory'' and subsequent
developments have shown that it is at least as fundamental to the overall
picture as the superstring theories, and possibly more so.  Little is known
about its definition beyond the low energy limit; we return to this subject
in the conclusions.

\xsubsection{Type \IIb}
Type \IIb\ superstring theory has a string with $T_1\sim 1/g_s\alpha'$
and a three-brane with $T_3\sim 1/g_s{\alpha'}^2$.  The strong coupling limit
is controlled by the string and must be another superstring theory
with chiral $\CN=2$ supersymmetry.  Thus it is the \IIb\ superstring --
this theory is self-dual.
The appropriate transformation of the
Lagrangian exchanges $B^{(2)}$ and $C^{(2)}$, and this requires
both an inversion $e^\phi\rightarrow e^{-\phi}$ of the coupling constant,
and a metric substitution (\ref{Weyl}).
The latter can also be described intuitively
as the rescaling of $\ap$ which when
combined with $g\rightarrow 1/g$ exchanges the two string tensions,
fundamental string $T_1={1\over\alpha'}$ with
solitonic string $T'_1={1\over g_s\alpha'}$.

This theory is particularly interesting as it is in fact symmetric
under general $SL(2,\BZ)$ transformations on the doublet $(B^{(2)},C^{(2)})$
and on the complex coupling $\tau=ie^{-\phi} + C^{(0)}$.
The duality described so far is the generator
$S:\tau\rightarrow -1/\tau$ and is usually called
``S-duality'' for this reason.  An orbit of $SL(2,\BZ)$
contains a spectrum of dyonic
strings with all $(B^{(2)},C^{(2)})$
charges $(p,q)$, $p$ and $q$ relatively prime. \cite{SchS}

This symmetry has a pretty eleven-dimensional origin  --
to see it, we need to explain T-duality, which relates the two type \II\
strings.  This is only present after compactifying a dimension on a circle,
so let us consider type \IIa\ on $\BR^9\times S^1$.
When the radius of the $S^1$ goes below the string scale, the lightest
state in the theory will be a string wound about the $S^1$, producing
a particle with mass $m=2\pi R/\alpha'$.  In fact there is a spectrum
of multiply wound strings with all masses $2\pi R n/\alpha'$, and
just as in our discussion of the strong coupling limit of \IIa, the
natural interpretation is that a {\it new} tenth dimension has appeared,
of radius $R'=\alpha'/2\pi R$.  The new theory is a type \II\ superstring
with the same fundamental string, but a detailed treatment shows that
introducing the new dimension flips the chirality conditions
and turns \IIa\ into \IIb. \cite{Schwarz}

Thus, \IIb\ in nine dimensions is T-dual to \IIa\ in in nine dimensions,
and can be obtained by compactifying M theory on a
$\BR^9\times T^2$ with both radii small.
Then the $SL(2,\BZ)$ action on \IIb\ is just large diffeomorphisms
of the $T^2$ \cite{SchS,Asp}.

\xsubsection{Heterotic}
The heterotic string theories admit the same reasoning,
and it leads to the conclusion that they
are dual to theories of five-branes.
This is true in a sense,
but it turns out that the five-branes are not the fundamental degrees
of freedom of the dual theories.

The reason that we should not immediately
accept the argument which worked so
well for type \II\ strings is that in $\CN=1$, $d=10$ theories,
the fundamental states need not be BPS solitons,
as shown by the example of the type \I\ string.

A little experimentation shows that
the S-duality transformation $\phi\rightarrow -\phi$ can be extended to a
duality transformation on all the massless fields, mapping the Lagrangian
(\ref{het}) into (\ref{typeI}).  This leads to the conjecture that
the $SO(32)$ heterotic string is dual to the type \I\ string.
The heterotic five-brane will then be dual to a solitonic
object, the five-brane solution of the
type \I\ effective Lagrangian.

This duality will be easier to establish
in the reverse direction, since the fundamental heterotic string
is a BPS soliton.

\xsubsection{Type \I}
The type \I\ string theory has $C^{(2)}$, supporting a BPS string
soliton and dual five-brane.
A strong test of the conjectured duality is that this string
must have the zero modes of the $SO(32)$ heterotic string.
It is quite a task to show that the string soliton
solution of the field theory indeed has these zero modes, and this
is where the Dirichlet brane comes to the rescue, as we will describe
in the next section.

An even stronger test is that heterotic string theory compactified on
a torus gains additional non-abelian gauge symmetry for particular
$O(\alpha')$ values of
the parameters (torus metric and flat gauge connection),
in a way not predicted by the low energy Lagrangian (\ref{het}).
What makes this test particularly interesting is that this happens
for values of the heterotic string parameters corresponding to a weakly
coupled dual type \I\ theory, so there is no obvious mechanism for
BPS states to become massless and we seem to have a contradiction.
This paradox and its very interesting resolution are described in
\cite{PolWit}.

These considerations
also lead to a proposal for the strong coupling limit of the $E_8\times E_8$
heterotic string -- it is M theory compactified on an interval \cite{HorWit}.
After compactification on $S^1$,
the two heterotic strings are continuously connected (using T-duality),
so this establishes a complete chain of dualities connecting all the theories.

%\end{itemize}

\section{Dirichlet branes}

A Dirichlet (D) brane can be defined in type \I\ and \II\ superstrings.
It is simply a hypersurface in space-time
on which open strings are allowed to end.  \cite{CJP}
The open strings are quantized just as in type \I\ theory,
with the difference that
the end points satisfy Dirichlet boundary conditions
$X^\mu(0)=X^\mu(\pi)=x^\mu$ for the coordinates normal to the hypersurface.
This makes sense for a hypersurface of any number of dimensions, say
$p$ space and one time, but consistency ultimately restricts $p$ to one
of the values for which the form $C^{(p+1)}$ or its dual $C^{(7-p)}$
appears in the Lagrangians
(\ref{typeII}) or (\ref{typeI}).

Just as quantizing open superstrings in ten dimensions leads to
ten-dimensional super Yang-Mills theory, the new open string sector also
contains a supersymmetric gauge theory as its massless sector.
The vector $A_\mu(x)$ is reduced to a $p+1$-component vector, while its
other components are replaced by scalar fields $X^i$ describing
fluctuations of the original hypersurface.
In the low energy, low amplitude
limit, the Lagrangian is the $d=10$ super-Maxwell
Lagrangian (a free theory) with
the substitution $A_i\rightarrow X^i/\ap$.

The identification of D-branes with the BPS RR charged $p$-branes rests on
their tension, charge and the fact that they preserve
half of the ten-dimensional supersymmetry \cite{Pol}.
The last statement is the simplest to see as it generalizes a
statement we made for the type \I\ string:
the boundary conditions of an
open string relate the two world-sheet spinors $\theta^1$ and $\theta^2$,
but the zero modes of a linear combination of the two remain.
Similarly, the relation $T_p \sim 1/g_s$ generalizes the $1/g_s=e^{-\phi}$
prefactor of the type \I\ open string action (\ref{typeI}).

Comparing to the $p$-brane discussion,
the fields $X^i$ describe small variations of the embedding $f$ of
(\ref{braneact}).
The additional world-volume gauge theory is new
but could have been predicted
from considerations of supersymmetry.
The real novelty of the D-brane appears when several parallel
branes are brought
into contact.  Now open strings stretching from one D-brane to another
produce new states, becoming massless when the D-branes coincide.

The open strings can again be distinguished by Chan-Paton
indices at each end, say $1\le i\le N$, each labelling a choice of D-brane.
The discussion is exactly as for type \I\ and
indeed type \I\ open strings can be regarded as associated with
$32$ D$9$-branes filling space-time.
The open string fields again become matrices, and are
governed by dimensionally reduced SYM with gauge group $U(N)$.
The Lagrangian is
\begin{equation}\label{Dlag}
\CL = {1\over g_s}\Tr F^2 + {1\over g_s\ap^2}\Tr (DX)^2
+ \bar\psi\Dslash\psi + {1\over g_s\ap^4}\sum \Tr[X^i,X^j]^2 .
\end{equation}
(For $p=3$ this is exactly the $\CN=4$ SYM of section 2, and we are
giving a different picture for the physics reviewed there.)
Separating the branes in space corresponds to giving a vev to the
matrix $X^i$.  The moduli space of configurations with zero energy
is $\BR^{N(9-p)}/S_N$, the space of diagonal constant matrices
$X^i_{mn}=x^i_n\delta_{mn}$ (solutions of $[X^i,X^j]=0$ modulo
gauge transformations).
$x_n$ is the position of the $n$'th D-brane, and
if all $x_n$ are distinct, the gauge symmetry is broken to $U(1)^N$.
By the standard analysis of the Higgs effect, off-diagonal states
$A_{\mu~mn}$ and $X^i_{mn}$ will get masses
\begin{equation}\label{masses}
m^2={1\over\ap^2}|x_{m}-x_{n}|^2.
\end{equation}
In other words, an open string stretched from brane $m$ to $n$ gains a
mass equal to the distance $|x_{m}-x_{n}|$ multiplied by the string
tension $1/\ap$.

Perhaps the simplest imaginable physical application of this is
to propose that the observable universe is a set of three-branes embedded in
ten-dimensional space, and that we have just described the origin of the
gauge symmetry of the Standard Model.  The breaking of gauge symmetry
is then associated with separating individual three-branes in the other
dimensions.

A potential problem with this idea is that the gravitational interaction
will not in general be described by four-dimensional
general relativity, as it is a `bulk' ten-dimensional phenomenon.
This problem can be solved if the additional dimensions are a small
compact space.  To judge how small is small enough, one must keep
all Kaluza-Klein gravitational modes in the effective four-dimensional
theory and compare their effects with the experimental bounds on deviations
from general relativity.  These bounds are weak and allow
sizes on the order of a millimeter!\footnote{
So far, similar models
which have been studied in more detail
(for example in \cite{Banks}) have other dynamical constraints
which keep this size microscopic,
but we do not know a general argument requiring this.}

\xsubsection{Remarks on the relation to noncommutative geometry}
The promotion of space-time coordinates to matrices certainly sounds like
it deserves the name ``noncommutative geometry,'' and indeed
the picture we just described
has a noteworthy similarity to Connes' construction of Yang-Mills theories
in terms of the noncommutative geometry of a discrete bundle
over space-time.  \cite{Connes}
In both pictures, the underlying space-time is a product
of the original space-time with
a set of points labelled by a ``fundamental index,''
so points correspond to pairs $(x,m)$.
In both pictures, gauge symmetry breaking is associated with
a non-zero distance between the points $(x,m)$ and $(x,n)$.

Making this correspondance more precise reveals some differences.
One is that the D-branes are
embedded in a higher-dimensional space-time, so the full theory
must contain more degrees of freedom.
A more important difference is the relation between distance and gauge
symmetry breaking.
In physical terms,
the Higgs fields and gauge symmetry breaking
of \cite{Connes} are exactly what would be produced by
straightforward dimensional reduction of a higher dimensional gauge theory,
$\phi_i=A_i$ with all fields taken independent of the additional
coordinates $X$.
A goal of \cite{Connes} is to define geometric concepts such as distance
in terms of natural operators, in this case the covariant derivative $D$
acting on functions $f_m(x)$.
The distance associated to this operator is the
difference $|f_m(x)-f_n(y)|$ maximized over all functions with $|Df|\le 1$.
Evaluating this for two points $(x,m)$ and $(y,n)$ with $x=y$
produces the distance $d(m,n) = 1/\phi_{mn}$.
In other words, the Higgs has units of inverse length.

Although this agrees with the usual dimensional analysis of field theory, it
is not the relation (\ref{masses}) of the D-brane construction.
In string theory, the presence of a fundamental constant with units
of $({\rm length})^2$ allows the scale of gauge symmetry breaking
to be proportional to a distance $X_{mm}-X_{nn}$,
and zero distance to correspond to enhanced gauge symmetry.

The meaning of this difference is made clearer by considering
compactification on a torus, after which
the dimensional reduction origin for scalars $\phi$ and the D-brane
origin for scalars $X$ are related by T-duality \cite{CJP}.
One way to say this is that in this case
the D-brane open string theories become secretly
ten-dimensional once all the degrees of freedom are included.
The additional degrees of freedom which appear on the torus are open
strings which start on D-brane $m$, wind any number of times $w^i$,
and end on D-brane $n$.  A field describing all of these modes would
be $A_{mn}(x,w)$.  This is Fourier dual to $A_{mn}(x,\tilde X)$ on a
dual torus and the statement of T-duality is that
these degrees of freedom are local fields on
the dual torus, with radius $\alpha'/R$.  Furthermore, displacing
one of the original D-branes by varying $X^i_{mm}(x,0)$ is T-dual to
varying the background flat connection on the dual torus,
$\phi^i_{mm}=\int d\tilde X^i\ A_{i,mm}(x,\tilde X)$.
Reversing the T-duality,
Higgs fields $\phi^i_{mm}$ originating by dimensional reduction
in string theory are proportional to a physical distance
on the dual torus.

%Of course there should be many ways to apply the mathematics of
%noncommutative geometry, and we have every expectation that it will be
%useful in describing D-brane and related physics.

\xsubsection{Physics of coincident branes}

The idea that when two solitonic objects approach each other,
extended objects stretched between them become light, leading
to new massless degrees of freedom when they coincide,
is clearly very general.  Indeed, dualities can be used
to relate every object in superstring theory to a Dirichlet brane (at
least in some limit of moduli space), showing that this is generic.

One role of these degrees of freedom is to produce bound states
of several D-branes.  String duality predicts additional BPS solitons
with non-unit charge, for example the dyonic strings of type \IIb\ theory.
These have been shown to exist as bound states of
D-strings and fundamental strings \cite{Wit3}.

The most striking and important application
of this so far is due to Strominger and Vafa
\cite{StromVaf}, and extended by many workers.
They considered D-brane bound states which are continuously connected (by
varying the vacuum moduli) to extremal black holes with event horizons.
These are BPS, which guarantees that the number of states of a given charge is
the same for the two systems.  It can be calculated in the
weak coupling limit of the D-brane system, and
thus the Bekenstein-Hawking entropy of extremal black holes can be calculated
from first principles in string theory.
This direction is advancing rapidly and any list of references would be
out of date by the time this reaches print.

Although the matrix nature of the D-brane coordinates $X^i$ is important
in describing the full dynamics, the moduli space of D-branes in flat space
was the subspace $[X^i,X^j]=0$.  At low energies where the moduli space
approximation is good, non-commutativity plays no role, and the embedding
into the full non-commutative configuration space is somewhat trivial.

However there are other geometries in which the non-commutativity plays a
more essential role, and the moduli space is embedded non-trivially in the
full configuration space.
An example is the description of D-branes
in an ALE space found in \cite{dm}.
ALE spaces are four real dimensional spaces with self-dual metrics
asymptotic to the flat metric on $\BR^4/\Gamma$, with $\Gamma$ a discrete
subgroup of $SU(2)$.  Self-dual metrics are hyperk\"ahler and these
metrics were constructed by Kronheimer as hyperk\"ahler quotients \cite{Kron}.
Now the moduli spaces of supersymmetric gauge theories (with $8$ supercharges,
as in the present case) are (by definition)
hyperk\"ahler quotients \cite{HKLR,Hitch}, and so an ALE space
can be obtained as the moduli space of a known SYM Lagrangian.
In ref. \cite{dm} it was shown that, starting with $\BR^4$ and an
action of $\Gamma$, the natural definition of D-branes on the quotient
is just this Lagrangian.  One introduces `image' D-branes on whose
Chan-Paton index the
regular representation of $\Gamma$ acts, and takes the
subsector of (\ref{Dlag}) invariant under the simultaneous action of $\Gamma$
on $\BR^4$ and this index.
The connection with the previous examples is that the construction relies
on massless degrees of freedom which appear when a D-brane coincides
with its images.

More explicitly, we let $g\in\Gamma$ act on the fields of
(\ref{Dlag}) with $N=|G|$ as $A_\mu\rightarrow r(g) A_\mu r(g)^{-1}$ and
$X^i \rightarrow R^i_j(g) r(g) X^j r(g)^{-1}$.
$r(g)$ is the regular representation of $\Gamma$ and
$R^i_j(g)$ a unitary representation in $SU(2)\subset SO(4)$.
The $\Gamma$-invariant subsector has gauge group the unitary group
of $L^2(\Gamma)$, and the moduli space of gauge equivalence classes of vacua is
$\BC^2/\Gamma$.

The string theory contains additional fields $\zeta^a$
which naturally couple to the Lagrangian (\ref{Dlag}),
and deform the relation $[X^i,X^j]=0$ to
\begin{equation}\label{ALEcon}
[X^i,X^j]= \zeta^a \sigma_a^{ij}
\end{equation}
(where $\sigma_a^{ij}$ are Pauli matrices generating the Lie algebra of the
commuting $SU(2)'\subset SO(4)$).
The $[X^i,X^j]$
are exactly the moment maps of the hyperk\"ahler quotient construction,
and after this deformation the moduli space is smooth with an ALE metric.

Thus the position degrees of freedom of D-branes on a manifold of
non-trivial topology and geometry are
embedded as non-commuting variables in a larger configuration space
of open string degrees of freedom.  We will return to the physics of this
in the next section.

\xsubsection{Non-identical branes}

So far we discussed parallel identical D-branes, but one could take D-branes
of different $p$, or non-parallel D-branes, and
stretched strings between them with mass $m^2 = m_0^2 + |\Delta X|^2$ appear
in every case.  The physics depends on $m_0^2$, the minimal $({\rm mass})^2$
for a bosonic string (fermionic strings always have $m_0=0$).
Generically $m_0^2<0$ and at small $\Delta X$ the theory becomes unstable.
The physics of this is quite interesting --
for example the annihilation of a $p$-brane
with an anti-$p$-brane begins with this instability \cite{BankSuss}
-- but still rather mysterious.

Certain geometries preserve some supersymmetry, which guarantees that
$m_0^2\ge 0$.
One example is the D$1$-brane of type \I\ theory, which implicitly is contained
in its D$9$-branes. This is the heterotic string soliton
required by type \I--heterotic duality.
Its additional degrees of freedom are strings stretching from $1$-brane
to $9$-brane; they will have a single $SO(32)$ Chan-Paton index, so they
must be the origin of the heterotic string fermions $\lambda^I$ of section 3.
Indeed one can check that $m_0^2>0$ for bosons and
the only new massless states are these chiral fermions.

The D$5$-brane of type \I\ theory also has strings stretched to the $9$-branes;
now $m_0^2=0$ so there are
additional massless charged scalars and a non-trivial moduli space.
Our earlier identification of the five-brane with a self-dual solution
of the Yang-Mills equations leads to an interesting result -- this
moduli space must be isomorphic to the moduli space of instantons.
It turns out that
this description is exactly the ADHM construction of moduli space
\cite{Wit2}.
Furthermore, the action of the D$1$-brane heterotic string soliton
in the presence of the five-brane should be equivalent to the sigma
model action for strings coupled to the self-dual solution, and making
this explicit reproduces the ADHM construction of self-dual gauge connections
\cite{Doug}.

\section{Short Distances in Superstring theory}

A long-standing belief of string theorists was that the finite size of
the string determined a minimum distance, below which the standard concepts
of space-time and metric would break down.  Many precise forms of this
statement were made.

One argument for this observes that
gravitons, fluctuations of the metric, are particular modes of the string.
At scales shorter than the string scale, there is no preferred way
to separate these modes from the other modes of the string and define a
unique space-time metric.
This is a real problem with the idea that metric could be an appropriate
description of small-scale geometry, and we will return to it below.

In studying structure at small scales, one should try to distinguish
properties of the background space-time from properties of the object one
uses to probe it.  Although in quantum gravity, any probe will affect the
background, clearly we should try to use the smallest and lightest
probe we can find.  However, the mass (or tension) of the probe controls
its own quantum fluctuations (it plays the role of $1/\hbar$ in
(\ref{braneact})) and this leads to an uncertainty relation: in quantum
gravity, the minimal length scale which can be resolved is the
Planck length \cite{DeWitt}.  In superstring theory, such reasoning leads to
the conjecture
that the minimal length scale should be the ten-dimensional Planck length as
defined in the Einstein metric, $1/l_{pl10}^8 = 1/\ap^4 g_s^2$.
To resolve it, we need a probe smaller than the fundamental string.

Dirichlet branes have a sort of intermediate status between fundamental
states and conventional solitons, because of their anomalously small
charge and tension $T \sim 1/g_s$.
Since Newton's constant is
$1/G_N \sim 1/g_s^2$, the gravitational (and other) fields around a
Dirichlet $p$-brane behave as $g_{ij} \sim g_s/r^{7-p}$.
As stressed by Shenker \cite{Shenker}, in the weak coupling limit,
the size of the brane, identified as the size of the region in which
the fields become strong, shrinks to zero as $l_p \sim g_s^{1/(7-p)}\sqap$.
For the natural probes ($p=0$ in \IIa\ or $p=1$ in \IIb\ and type \I),
this is larger than $l_{pl10}$ but smaller than $\sqap$.

However, because the derivatives of the fields become large in this region,
using the low energy Lagrangian to make this argument is not valid.
To correct the argument, we must consider the details of how
these fields and interactions are realized in string theory.

When we bring two branes near each other, they will interact.
In the original field theory, the two branes are described
by a complicated multisoliton solution.
In the limit of large separation, the interaction is mediated by the
long-range massless fields of (\ref{typeII}) or (\ref{typeI}).
Their leading long-range behavior depends only on the mass and charge of the
source, and a good approximation for the action of the combined system is
obtained by substituting this into the $p$-brane action (\ref{braneact}) of the
other brane.

The D-brane description of the interaction is very
different.  New open string degrees of freedom appear.
Even if we do not excite them directly,
in the quantum
theory they will have zero-point energies, which can depend on the
distance between the branes and the other parameters of the system.
This produces a sort of `Casimir effect' interaction between the branes.

Although these two descriptions of the interaction are
quite different, the same
diagrams in perturbative string theory are responsible for both
of them.
This is simplest to see in the D-brane description and at
leading order in the string coupling.  This
contribution is mediated by a string world-sheet with annulus topology,
and with its two boundaries constrained to sit on one or the other D-brane.

The diagram can be interpreted in field theory terms in two ways.
We can regard it as a sum of closed strings emitted by the first D-brane,
propagating and absorbed by the other.  This includes the classical
gravitational and abelian
gauge interactions, along with the exchange of an infinite
tower of massive modes of the closed string.
Alternatively,
we can regard it as a sum of loop contributions from open strings
stretched between the branes, the lightest having
$({\rm mass})^2 = m_0^2+(\Delta X)^2$,
but again including an infinite tower of stringy excitations.

World-sheet duality states that we do not add these two sums but rather
that they are two descriptions of the same physical
amplitude.  However, the closed string sum is a more physical description
when the separation between the branes is
much larger than than the string scale $\sqap$, while the open string sum
is more physical if the separation is much smaller.
This is because the leading behavior in either of the limits is entirely
produced by the leading term in the appropriate sum, the contribution of
the lightest states.
To a very good approximation,
the infinite towers of more massive states can be neglected,
returning us to a field theory description.  But the
relevant field theory is different in the two regimes.

Thus string theory incorporates both pictures of the interaction
between branes, and provides an interpolation between them.
Having seen this, we now know that at short distances,
we should not trust even the qualitative behavior
of the original supergravity solutions describing the branes.
The correct description of the short distance interaction between
branes is given by the supersymmetric gauge theory (\ref{Dlag})
and its generalizations.

\xsubsection{D$0$-brane scattering}

A good example which illustrates the resulting physics is the
scattering of two D$0$-branes.
This was studied semiclassically in \cite{Bachas}
and in terms of open string quantum mechanics in \cite{DFS,kp}.
It was shown in \cite{DKPS} that each of these approximations is
controlled in a specific physical regime, leading to definite results
for energies below the D$0$-brane mass, and thus probing length scales
$l > g_s\sqap$.

Gravitational scattering of point
particles is always `hard' with a significant amplitude for large
deflection, just like the Rutherford scattering of electrons
off nuclei, and for the same reason -- the interaction grows at short
distance like an inverse power of $r$.  However, D$0$-branes scatter
differently, because the gravitational interactions are not fundamental
but are derived from the quantum fluctuations of open strings.
For any fixed total energy $E$ and corresponding velocity $v$
(with $E=v^2/2g_s$), there exists a distance $b\sim \sqrt{v}$ such
that at $r<b$ the open string effects cannot track the motion of
the D-branes (the quantum mechanical Born-Oppenheimer approximation
breaks down) and the force law is modified.
This leads to a softer
scattering, controlled (surprisingly from the ten-dimensional
string theory point of view) by the momentum measured in units of the
eleven-dimensional Planck length.  Indeed the importance of this scale in
D$0$-brane physics was predicted by a simple scaling argument in \cite{kp}.

One can show \cite{DKPS} that in this problem the quantum open string effects
also reproduce the long range fields; in particular the motion of the
D-branes at long distances is described geometrically, as one D-brane
moving in the metric produced by the other.  The fact that the gauge
theory description could reproduce this exactly is special to this and
a few other problems with extended supersymmetry, but in general there
must be a geometrical description of the motion interpolating between
the known long-distance metric
and a short-distance metric computable from
the D-brane gauge theory -- essentially, the natural metric on the moduli space
of ground states of the gauge theory.
This identification of the short-distance metric as a derived quantity in the
D-brane gauge theory
escapes the paradox mentioned at the
start of the section, that the original graviton mode of the string field
has no sensible definition at sub-stringy scales.

An important lesson of the scattering exercise is that the
geometrical description breaks down at very short distances.
The lower the
energy of our probes, the shorter the distance scale of the breakdown,
but such a breakdown will always occur whenever the original geometry
was singular.  This is a special case of a general principle in
field theory: singularities in effective field theory are produced by
integrating out states which become massless, i.e. keeping only the
functional dependence of their quantum effects on the other degrees of
freedom.  By keeping their full dynamics, one obtains a non-singular
description.  The effective field theory here governs the motion of the
D-branes at low velocity; the states we integrate out are the open strings
stretched between the D-branes; they become massless when the D-branes
coincide, and by keeping them in and treating the full theory (\ref{Dlag})
we get the true, non-singular dynamics.

To illustrate how these concepts apply in a situation with non-trivial
topology, we reconsider the description of ALE space given
in section 4.2.  The remarks above motivate the claim
that this is the appropriate description whenever the volumes of
the homologically non-trivial two-cycles in the space are much less than
the string scale, ${\rm Vol}(S^2) << \ap$, while the normal metric and
manifold description is appropriate when ${\rm Vol}(S^2) >> \ap$.
The continuous connection between the two descriptions this implies
is still somewhat mysterious, though the interpolation
should be smooth.

For definiteness we consider a single D$0$-brane.  At low energies its
motion will be governed by its gauge theory moduli space metric, which is
just the ALE space.  However this is the submanifold defined by (\ref{ALEcon})
of a larger space, topologically
trivial but carrying a natural noncommutative algebraic structure.
At finite energies the dynamics can explore more of this configuration
space, and when the energy exceeds the height of a saddle point of the
gauge theory potential,
$E \sim ({\rm Vol}(S^2)/\ap)^2/g_s$,
the topology of the available configuration space changes.
In this regime the original space-time interpretation is completely
inadequate -- it sits in something more fundamental.
This will be true even at low energies
if ${\rm Vol}(S^2) \sim g_s^{2/3}\ap=l_{p11}^2$, as
quantum fluctuations will dominate the effects of the potential.
D$0$-brane probes indeed see a minimum length scale, but it is the
eleven-dimensional Planck length.

\medskip

To summarize,
space-time indeed has a geometric description at sub-stringy
scales.  This is space-time as seen by D-branes, but these are
the only objects we know in string theory which could resolve such scales.
The metric is no longer fundamental but is
instead derived from other fundamental degrees of freedom, the open strings
responsible for the dynamics of the D-brane.
Since ten-dimensional quantum gravity is not valid at short distances, the
ten-dimensional Planck length plays no role.  The eleven-dimensional
Planck length plays a surprisingly important role, perhaps pointing to
a more fundamental role for M theory.

The geometric picture of space, either around a brane or in non-trivial
topology, is an approximate description valid at low energies.  When D-branes
approach within short distances or encounter small scale structures,
new degrees of freedom become light, and the
correct description of the dynamics is
to treat the new degrees of freedom
on an equal footing with the original coordinates.  Thus the
original geometric description is naturally embedded in
a larger configuration space, which in these examples is a matrix
generalization of the algebra of space-time coordinates.

\section{Further Directions}

In this talk we only described ten and eleven-dimensional theories.
Much of the beauty and interest of superstring duality appears after
compactification to lower dimensions.
This is a vast subject and we will confine ourselves to the single
remark that the results motivate the point of view that all
of the $p$ and D-branes we have discussed are in some deep sense the
same \cite{Town}.
The evidence is that for every soliton, there is some limit
in which it plays a fundamental role.  For example, a $p$-brane gives
rise to a string after compactification on $T^{p-1}$, by wrapping it
around the $T^{p-1}$ (i.e. taking its hyperplane to be $T^{p-1}\times \BR^2$).
The lower dimensional theory then has a discrete duality
symmetry which exchanges this object with the fundamental string.

Much remains to be understood about short distance structure in string theory.
One would conjecture that every local structure with non-trivial topology has
a specific small scale realization as an embedding in a larger configuration
space.  How this description matches on to normal geometry and topology and
the role of general covariance in the complete picture are quite mysterious.
It is unclear whether supersymmetry plays an essential or merely facilitating
role in the discussion.  The treatment of \cite{DKPS}
allows studying shorter lengths than $l_{p11}$ but is valid only up to
energy scales $E \sim 1/g_s$ at which pair production of D-branes becomes
possible; it is unclear whether space-time is sensible at smaller scales or
what should replace it.

Eleven-dimensional M theory is at the very least a new regime of the
single unified theory with quite different properties than string theory,
and very possibly more fundamental.
%The Lagrangian
%(\ref{sgeleven}) is certainly the simplest of the four.
%
M theory contains no fundamental string but instead a
membrane and five-brane.  The \IIa\
string arises as the membrane wrapped around the $S^1$.
One naturally wonders
if the membrane can be quantized and used as the fundamental object.
It should be realized that
unlike quantizing the superstring, this is not a problem
of quantizing a known classical Lagrangian.  The membrane
world-volume theory is non-renormalizable and its physics depends strongly on
the specific cutoff used to define it.
This is in contrast to the string whose renormalizability implies
that independent of their short-distance definition, all superstrings will
have the same long distance physics as the fundamental string.

Direct attempts at membrane quantization have not yet succeeded.
Perhaps the D$0$-branes of \IIa\ theory are closer to the fundamental
M theory degrees of freedom, as in the work \cite{BFSS} (which also finds
interesting connections with noncommutative geometry).

\xsubsection{A complete formulation of string/M/F/...-theory}

This talk was addressed to both physicists and mathematicians, and
perhaps for the mathematicians, there is another question
which would seem to have priority over
the two I listed in the introduction: what are we talking about?
In other words, can we define the theory.
Given that this is hard even for field theory (as we mentioned in section 2,
giving a definition of $\CN=4$, $d=4$ SYM for which duality can be precisely
formulated would be a real advance) I will only make some general comments.

Let me first comment that there is striking evidence that we have
all the degrees of freedom.  The Bekenstein-Hawking entropy of a black
hole comes out of very general principles of general relativity,
quantum field theory and thermodynamics, with no dynamical input.
Strominger and Vafa's success in reproducing this from microscopic
dynamics, and subsequent success in extending this to non-BPS black holes,
strongly suggests that there are no hidden degrees of freedom waiting
to be discovered.  If so, we have some solid ground from which to start.

In the light of duality, we should start by
re-examining just what we mean by a complete definition.
Generally,
we start off with a list of physical properties we believe
our theory possesses -- a consistent probability interpretation (for
quantum theories), locality and causality, and a symmetry group.
Usually a definition will not make all of these properties
manifest and we must prove the others.

Usually quantum field theory is based on a set of local
fundamental degrees of freedom,
which become independent in some limit.
But, in a theory with duality, it is not natural to
single out a particular set of
degrees of freedom as fundamental.  When we do, the physics of
the other limits becomes inaccessible.

One might try to make a precise definition
for each possible choice of fundamental object, show that these have
overlapping regimes of validity, and
and find some sort of explicit change of variables or `transition function'
on the overlaps \cite{Vafa}.  This is of course just what we have
been doing, but the variables have not been defined precisely, and the
change of variables is only known explicitly for an abelian subsector of the
theory, non-interacting at low energies.

The ideal definition of string/M/F/...-theory would
have a sort of `manifest duality' in which every
object which could become fundamental in any limit was included as
a fundamental degree of freedom.
However, there is a bewildering variety
of candidate fundamental degrees of freedom,
especially after compactification, and such a description might well
require intractable constraints among them.

One might look for some general `principle of construction' that builds
up all possible extended objects as composite objects.
Perhaps the appropriate constituent degrees of freedom
have not even made their appearance yet!

The assumption of local fundamental degrees of freedom
is not logically necessary,
and more general frameworks have been studied, only imposing locality
on the observables.  Indeed locality has never been well understood
even in perturbative string theory.  For example, there exists no formulation
of the state at a specified time in string theory, so the initial value
problem cannot even be properly stated, much less shown to be causal.

These are all indications that a fully satisfactory formulation of superstring
theory or M theory is still far beyond us.  At least we now have some
feeling for the non-perturbative physics it is supposed to represent.
New insights from both physicists and mathematicians will surely be
required before we have it.

\bigskip
I would like to thank the organizers and participants
for a stimulating school, the Rutgers high energy theory group and its many
1995-96 visitors for teaching me about many of these topics and for
enjoyable collaboration, and Scott Axelrod, Alain Connes and Is Singer
for giving me some idea of a mathematician's point of view on these topics.

\end{document}